\documentclass[useAMS,usenatbib]{mn2e}

\usepackage{graphicx,amssymb,bm}
\usepackage{subfig,xfrac}
\usepackage{float, Abbrev}
\usepackage[fleqn]{amsmath}  
\usepackage{color}

\newcommand{\be}{\begin{equation}}
\newcommand{\ee}{\end{equation}}
\newcommand{\ba}{\begin{eqnarray}}
\newcommand{\ea}{\end{eqnarray}}

\def\simlt{\lower.5ex\hbox{$\; \buildrel < \over \sim \;$}}
\def\simgt{\lower.5ex\hbox{$\; \buildrel > \over \sim \;$}}

\usepackage{graphicx} 

\title[A weak lensing comparability study of galaxy mergers that host AGNs]{A weak lensing comparability study of galaxy mergers \\ that host AGNs}
\author[D. Harvey \& F. Courbin]
{D. Harvey$^{1}$\thanks{e-mail: {\tt david.harvey@epfl.ch}} and F. Courbin$^{1}$ \\
$^{1}$Laboratoire d'astrophysique, EPFL, Observatoire de Sauverny, 1290 Versoix, Switzerland}

\begin{document}

\date{Accepted ---. Received ---; in original form \today.}

\pagerange{\pageref{firstpage}--\pageref{lastpage}} \pubyear{2013}

\maketitle

\label{firstpage}

\begin{abstract}
\noindent 
We compared the total mass density profiles of three different types of galaxies using weak gravitational lensing: 
(i) 29 galaxies that host quasars at $\bar{z}\sim0.32$ that are in a post-starburst (PSQ) phase with high star formation indicating recent merger activity, 
(ii) 22 large elliptical galaxies from the SLACS sample that do not host a quasar at $\bar{z}\sim0.23$, 
and (iii) 17 galaxies that host moderately luminous quasars at $\bar{z}\sim0.36$ powered by disk instabilities, but with no intense star formation. 
On an initial test we found no evidence for a connection between the merger state of a galaxy and the profile of the halo, with the PSQ profile comparable to that of the other two samples and consistent with the \cite{cosmos_quasar} study of moderately luminous quasars in COSMOS.
Given the compatibility of the two quasar samples, we combined these and found no evidence for any connection between black hole activity and the dark matter halo. 
All three mass profiles remained compatible with isothermality given the present data.

%
%
%
%
%
%
%
%
%
\end{abstract}


\section{Introduction} \label{sec:intro}
The mass of a black hole (BH) has been shown to tightly correlate with the size of the stellar bulge of its host galaxy and even tighter with its stellar velocity dispersion \citep{BH_bulge,BH_bulge2}, implying an inextricable link between galaxy evolution and BH formation  \citep{BH_gal_form,BH_gal_form2}.

Efforts to construct a single unifying theory that can explain the formation and fueling of BHs were set back with the advent of the Hubble Space Telescope (HST) \citep[e.g.,][]{Bahcall1997, disney95}.
With the improved angular resolution of a space-based telescope and with deep spectroscopy, BHs were found indiscriminately in old, early-type elliptical galaxies and younger, smaller, spiral galaxies \cite[e.g.][]{Letawe2008, Letawe2007, Jahnke2007}. 
As large elliptical galaxies with brighter central quasars are more common at higher redshifts than spiral galaxies with less luminous BHs, and vice versa at lower redshifts, it was suggested that the average cosmic quasar activity for a given galaxy dropped as the Universe evolved to lower redshift \citep[e.g][]{dunlop03,mclure99,downsize}.  

This ``Cosmic Downsizing" suggested that there existed multiple fueling mechanisms that could sustain BHs, leading to a two-mechanism paradigm in BH formation \citep{downsizing,downsizing1}. 
The first mechanism to produce a BH occurs earlier in the Universe where the merger of two galaxies can instigate an intense burst of star formation and strong inflows, triggering the ignition of a quasar. 
The quasar phenomenon, the luminous counterpart of a supermassive BH, is an important stage of a galaxy's life, since the resulting feedback blows out any local dust, quenching star formation and ultimately evolving the galaxy onto the red sequence \citep{merger_qso,merger_qso1,merger_qso2}. 
The second mechanism occurs in spiral galaxies at lower redshifts where a large central bar can contain sufficient gas to continually accrete and fuel a BH \citep{accretion_qso,accretion_qso1}. This mechanism results, however, in much less luminous quasars. 

In this letter we compared the effect these two mechanisms had on the density profile of quasar host galaxies using weak gravitational lensing. 
To study the former mechanism, i.e. that BHs are triggered by merging events, we looked at galaxies that showed recent signs of intense star formation, as it is thought that any major merger will instigate the formation of large quantities of stars. The HST snapshot survey of 29 Post-Starburst Quasar (PSQ) showed clear signs of recent intense star formation \citep{PSQ1,PSQ2} and is was suited to this study. This point, during or just after a merger could theoretically have produced a steeper profile as a result of dynamical friction \citep[e.g.,][]{Dobke2007, Auger2008}. We therefore directly probed the slope of the mass profile of quasar host galaxies to test whether quasars do in fact tend to form in steep-profile dynamically perturbed galaxies or in shallower-profile (isothermal) more relaxed galaxies.

We compared the mass density profile of PSQ host galaxies with two comparison samples. We first compared them to a sample of galaxies that did not harbor an active central black hole and a second that did but that did not show any evidence for recent merging activity. For the first comparison sample we considered 22 non-quasar early type galaxies from the Sloan Lens ACS Survey \citep[SLACS;][]{SLACS} for which weak and strong lensing measurements had been made \citep[][hereafter G07]{SLACS_SL}. The second was a sample of quasars from the Co-evolution of spheroids and black holes HST program \citep[][hereafter T07]{coevolutuionBH}. These moderately luminous Seyfert galaxies represented a typical set of quasars whose host galaxies did not have strong star formation but did contain an active galactic nuclei \citep{BHstarform}. By comparing the halo of the PSQs to the SLACS and T07 halos we examined any differences that may indicate a connection between the mergers that triggered a quasar and the mass profile of the quasar host galaxies.


Tracing the differential mass components of a galaxy has now become commonplace amongst large scale sky surveys. 
Exploiting a galaxy's gravitational influence on the local space-time allows a non-parametric and assumption-free analysis of the projected density profile. 
Weak gravitational lensing of galaxies by galaxies or `galaxy-galaxy lensing', describes the effect that a foreground galactic potential has on the observed shape of a background, `source galaxy'.
By measuring the shape of background galaxies and correlating them to foreground galaxies it is possible to extract the underlying projected density profile. The first detection of galaxy-galaxy lensing was by made by \cite{gal_gal}. This pioneering analysis studied the gravitational lensing signal around galaxies and attempted to fit a single component mass model to the data, however this model assumed all the mass was contained within the galaxy halo \citep{gal_gal_1halo,gal_gal_1halo1,gal_gal_1halo2}. Subsequent studies found that either analysing isolated galaxies \citep{gal_gal_isolated}, or applying more sophisticated models that included second halos were required \citep{gal_gal_2halo,gal_gal_2halo1,gal_gal_2halo2}.  One of the most recent study was from the CFHTLenS survey \citep{CHFTLenS}. Studying the relationship between the dark matter and stellar halo around a galaxy \citep{chft_gal_gal}, this ground based survey used galaxies at $0.2<z<0.4$ and applied a multi-component model to dissect the contributions from stars, dark matter and surrounding satellite halos. Although complimentary to the study here, the advantage of 155 square degree survey was that the analysis could extend out to 10~Mpc, whereas here we analysed the more central parts, in the range $10<r<1000$ kpc, where quasar feedback has most impact. 

Studies specifically probing the host galaxy of a quasar are few.
Most recently, the profile of galaxies hosting Active Galactic Nuclei (AGN) has been mapped using the COSMOS quasar catalogue \citep{cosmos_quasar}. At a mean redshift of $\bar{z}=0.7$, they used 382 moderately luminous X-ray AGN and found that quasars tended to lie in medium-sized halos with $M_{halo}\sim10^{12}M_\odot$, with a halo occupation distribution not too dissimilar from normal galaxies that do not host quasars. At this redshift the majority of the AGN are fueled by the galactic bar and therefore exhibit a similar profile to that of normal galaxies, which is indeed supported by their measurements. Prior to this, \cite{agn_gal_gal} studied the dark matter halo of radio loud quasars in the Sloan Digital Sky Survey and compared them to a sample of control galaxies finding that these AGN laid in larger dark matter halos than the galaxies. 
In this study we build upon and extend this body of work by directly probing the dark matter halo of galaxies that are in a specific period of their life, where discrepancies in the dark matter halo may be most evident.

In the following, we adopt a flat cosmology with $H=100 h^{-1}$km~s$^{-1}$Mpc$^{-1}$, $\Omega_{\rm M}=0.3$ and $\Omega_\Lambda=0.7$.

\section{Data \& Method}

The HST snapshot survey of 29 post-starburst quasars \citep[Proposal ID 10588;][]{PSQ1,PSQ2} was well suited to our study. The sample was imaged using  the HST F606W filter on the Advanced Camera for Survey (ACS) at a mean redshift of $\bar{z}=0.32$ . We compared these galaxies to two control samples built from archived HST/ACS programs. The first was composed of massive elliptical galaxies from the Sloan Lens ACS Survey (Proposal ID 10494 and 10886), for which 22 strongly lensed galaxies were imaged through the F814W filter of ACS \citep[][G07]{SLACS} at a mean redshift of $\bar{z}=0.23$. These galaxies had been shown to be representative of the general populations of elliptical galaxies \citep{Barnabe2011}. The second contained 17 Seyferts galaxies that hosted moderately luminous AGNs  at a mean redshift of $\bar{z}=0.36$ \citep[Proposal ID 10216][]{coevolutuionBH}. The host galaxies of these quasars showed no obvious sign of any merger with no significant active star formation. For detailed selection criteria we point the reader to the cited references. 

In all cases we first corrected the images for charge transfer inefficiency \citep{CTI,CTI2}, and we calibrated and drizzled each image individually for PSF estimation and then we combined them for our final shape measurement. We then used the RRG pipeline from the COSMOS weak lensing survey \citep{RRG,COSMOSintdisp} and developed by \cite{Harvey15}, to measure the shapes (ellipticities) of the faint galaxies in the field of view. Although we did not have photometric or spectroscopic redshifts of the background galaxies, the quasars were at sufficiently low redshift ($z<0.5$) that we could assume that all measured galaxies were in fact behind and at a redshift of $z=1.0$ \citep{COSMOSintdisp}. We initially made the same magnitude cuts as G07, removing galaxies that were outside the interval $20 < M < 26.5$ however we found that we obtained significant contamination from bright sources near the lens and therefore cut between $21 < M < 26.5$  in all three samples. We also cut in size removing any galaxies that are less than 3 pixels in size and a signal-to-noise of less than 4.4. We also removed pairs of galaxies that were within $0.5"$ of one another.

Having measured the ellipticities of the background galaxies, we azimuthally averaged the tangential ellipticity in radial bins around the foreground objects in the three above samples and we converted this to the projected surface density in a similar way to G07 via
\be
\Delta\Sigma = \frac{\sum^{\rm N_{\rm lens}}_{j=1}{\Sigma^{-1}_{{\rm crit},j}\sum^{N_{s,j}}_{i=1}{e_{{\rm t},i}\sigma^{-2}_{e,i}}}}
{\sum^{\rm Nlens}_{j=1}\Sigma^{-2}_{{\rm crit},j}\sum^{N_{s,j}}_{i=1}\sigma^{-2}_{e,i}},
\label{eqn:excess}
\ee
where $\Delta\Sigma$ is the average excess surface density in a given radial bin,
\be
\Sigma_{\rm crit}=\frac{c^2}{4\pi G}\frac{D_{\rm os}}{D_{\rm ls}D_{\rm ol}},
\label{eqn:crit}
\ee
$e_{\rm t}$ is the tangential ellipticity around the galaxy we are probing, and $\sigma_{\rm e}$ is the error in the ellipticity, and subscripts refer to the relative angular diameter distances between the observer(o), the source(s) and the lens(l). To calculate the value of $\Sigma_{\rm crit}$ for each lens we used the reported redshift and assume that the source is at a redshift of $z_{\rm source}=1.0$. The first summation in equation \eqref{eqn:excess} is over  $N$ galaxies that lie in the radial bin, and the second summation is over the sample of galaxy lenses, $N_{\rm lens}$.

\section{Results}

\begin{figure}
\begin{center}
       \includegraphics[width=0.5\textwidth]{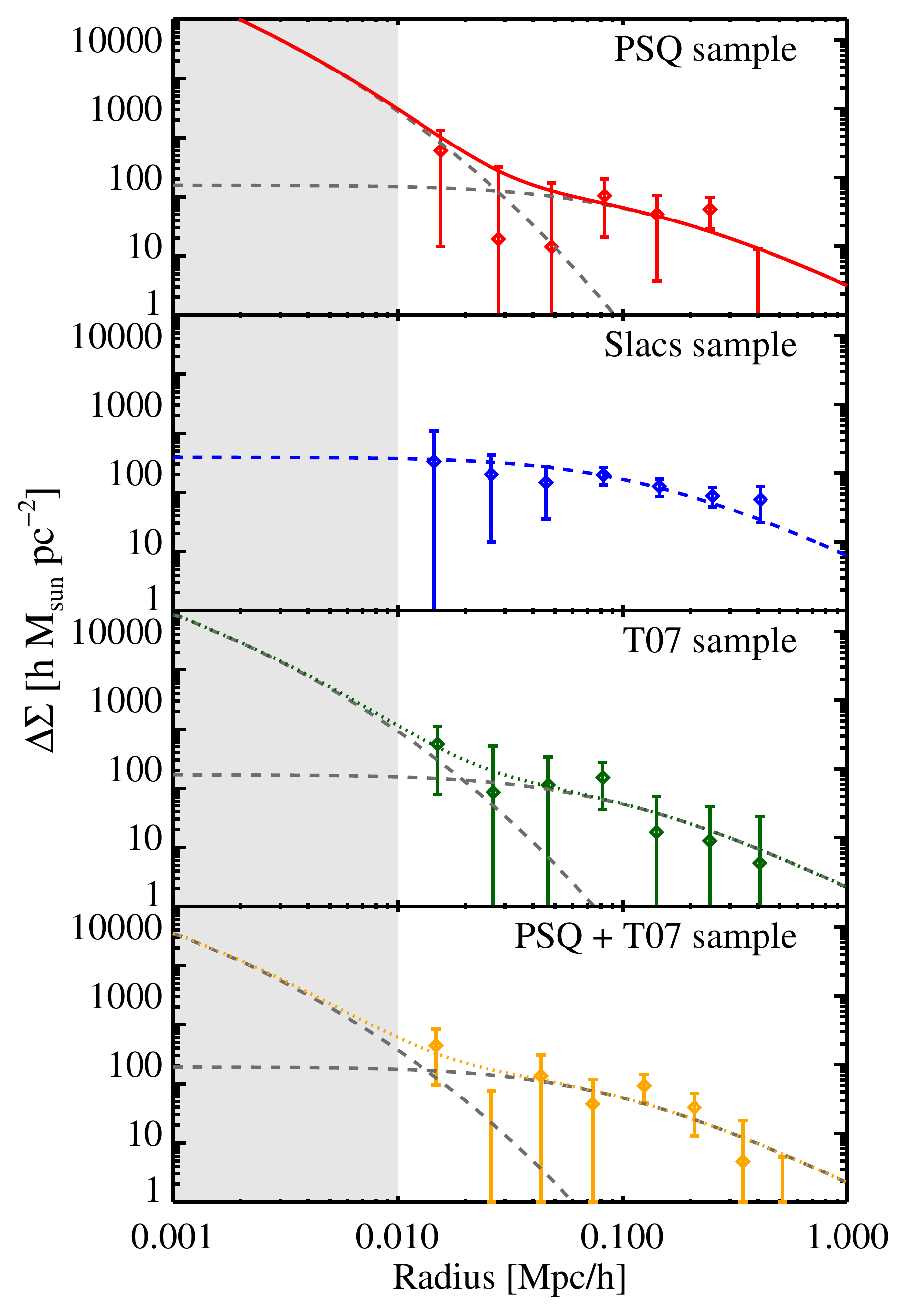}

       	\caption{Weak lensing signal for different types of galaxies. From top to bottom we show the excess projected average surface density for: a) the 28 host galaxies of PSQs, b)  22 massive early-type galaxies from SLACS with no active supermassive BHs at their centre, c) the 17 Seyfert galaxies of T07 and d) the combined quasar sample T07+PSQ. Note that for the SLACS sample we perform our own weak lensing analysis. Our measurements are consistent with the ones from G07. The shaded region denotes the regime that is too close to the lens to gain robust weak lensing constraints, and hence any fit is a direct extrapolation. }
	\label{fig:quasars_slacs}
\end{center}
\end{figure}

The top three panels of Figure \ref{fig:quasars_slacs} show our results for the different samples of galaxies, where errors shown are the 1$\sigma$ error on the mean. Given the small sample we jack-knifed each lens to ensure that no one object was biasing our results. We found the errors derived from this test were consistent with those shown here.  We fitted a Navarro, Frenk and White \citep[][NFW]{NFW} profile to describe the large-scale dark matter halo \citep{proj_NFW} and a de Vaucouleurs profile \citep{vauc,vauc1} to model the more central part of the total mass profile. 
Note that in doing so, we did not attempt to use any prior knowledge from the light distribution other than fixing the effective radius. This would require assumptions on the stellar populations of the galaxies, which were all very different in the three samples we use. 
We restricted ourselves in the present work to the total radial mass profile using galaxy-galaxy weak lensing. We fixed the effective radius of the de Vaucouleurs in each case using the reported results from the three main papers and hence only fitted the normalization of the de Vaucouleurs, and the normalisation and scale radius of the NFW halo. In each panel of Figure \ref{fig:quasars_slacs} we show the best fitting model and in grey the two contributing components. Table \ref{tab:results} gives the best fitting masses derived from the fits in units of $10^{12}h^{-1}M_\odot$, the derived NFW concentration and the chi-square of the best fit over the number of degrees of freedom. As a cross reference, we found that the NFW masses for the SLACS sample were consistent with the NFW masses in G07 analysis of $M_{\rm NFW}=14^{+6}_{-5} \times10^{12}h^{-1}M_\odot$.
\begin{table}
\begin{center}
\begin{tabular}{|c|c|c|c|c|}
\hline
Sample & $M_{\rm VAUC} $ & $M_{\rm NFW}$ & $c_{200}$& $\chi^2$/dof \\ 
\hline
PSQ      & $1.7\pm1.2$     &    $4.0\pm3.0$  & $3.6\pm2.8$  &$4.90/5$\\
SLACS   &     $ <0.84 $    &     $13.0\pm3.6$ &$5.4\pm1.5$ &    $4.35/5$\\
T07 & $0.72\pm0.8$     &    $4.4\pm3.2$ &$4.1\pm3.0$  &   $2.51/5 $  \\
PSQ + T07 & $0.18\pm0.57$ &  $2.8\pm1.9$ & $4.9\pm3.5$ & $5.75/5.$ \\
\hline
\end{tabular}
\caption{ The derived masses and concentration parameters from the best fitted models to the data. The first column stipulates the sample of galaxies, the second column gives the mass within the effective half light radius of the best fitting de Vaucouleurs profile in $10^{12} h^{-1}M_\odot$, the third column gives the best fitting NFW mass also in $10^{12}h^{-1}M_\odot$ and the fourth column gives the derived concentration with error. The final column gives goodness of fit to data for the reported model (chi-square / degrees of freedom).}
\label{tab:results}
\end{center}
\end{table}

The result derived from the two quasar samples implied they were compatible and drawn from a similar population. To test this conclusion we compared their velocity dispersions. Using the known relation between black hole mass and velocity dispersion from \cite{bhvelrel} we found that the PSQ sample had a mean velocity dispersion of $\langle\sigma_{\rm v}^{\rm PSQ}\rangle=208\pm41~$km/s and the T07 sample had a velocity dispersion of $\langle\sigma_{\rm v}^{\rm T07}\rangle=228\pm36~$km/s. Given the small sample size, we carried out a KS test to decide whether the two samples could have been drawn from two different populations and we found no significant evidence that they did ($P=0.35$). Given their similarity we combined the two sets of quasar host galaxies to test whether the resulting increased sample differed from the SLACS sample. 

The bottom panel of Figure \ref{fig:quasars_slacs} gives the combined profile of galaxies that contain an active supermassive black hole at its centre. We found that when combined there was no significant difference in the profile shape between these galaxies that did harbour an active central black hole and the profile of a galaxy that did not (SLACS). In addition, this profile did not significantly depart from that of an isothermal sphere. This result was consistent with \cite{cosmos_quasar} in which they found no difference in profile for galaxies containing moderately luminous quasars. However we did find a significantly different NFW mass, with the quasar sample lying in much smaller dark matter halos than the SLACS galaxies. We compared the velocity dispersions of the two quasar samples with the SLACS sample and found that the SLACS sample with $\langle\sigma_{\rm v}^{\rm SLACS}\rangle = 244 \pm 44~$km/s was consistent with the T07 sample (P=0.59) however, the distribution of velocity dispersion did differ from the PSQ by $\sim2\sigma$ (P=0.05). 
Our conclusion is that there is a difference between the PSQ host galaxies and the SLACS host galaxies, however this cannot be attributed to recent merging activity and is most likely due to the environment and formation history.

\subsection{Systematic Tests}

We verified our weak lensing measurements by carrying out two standard validation checks. We first checked the excess surface density derived from the cross-shear, i.e. the stretch along the axis $45^\circ$ to the tangential shear. Under the assumption there are no other external factors such as lens light impacting the shape measurement or poor PSF removal, the cross-shear should average out to zero, and hence the surface density calculated should also be zero. Our second test measured the tangential shear around random points in the field, which should also be zero since they are not correlated in any way. We found in both cases that all of the samples exhibited no significant excess signal. We also carried out a similar test but with the two components of ellipticity: $e1$ and $e2$, where in the absence of any additive bias the mean of these will be equal to zero. We check all three samples and we found no evidence for any significant bias.
\begin{figure}
\begin{center}
\includegraphics[width=0.47\textwidth]{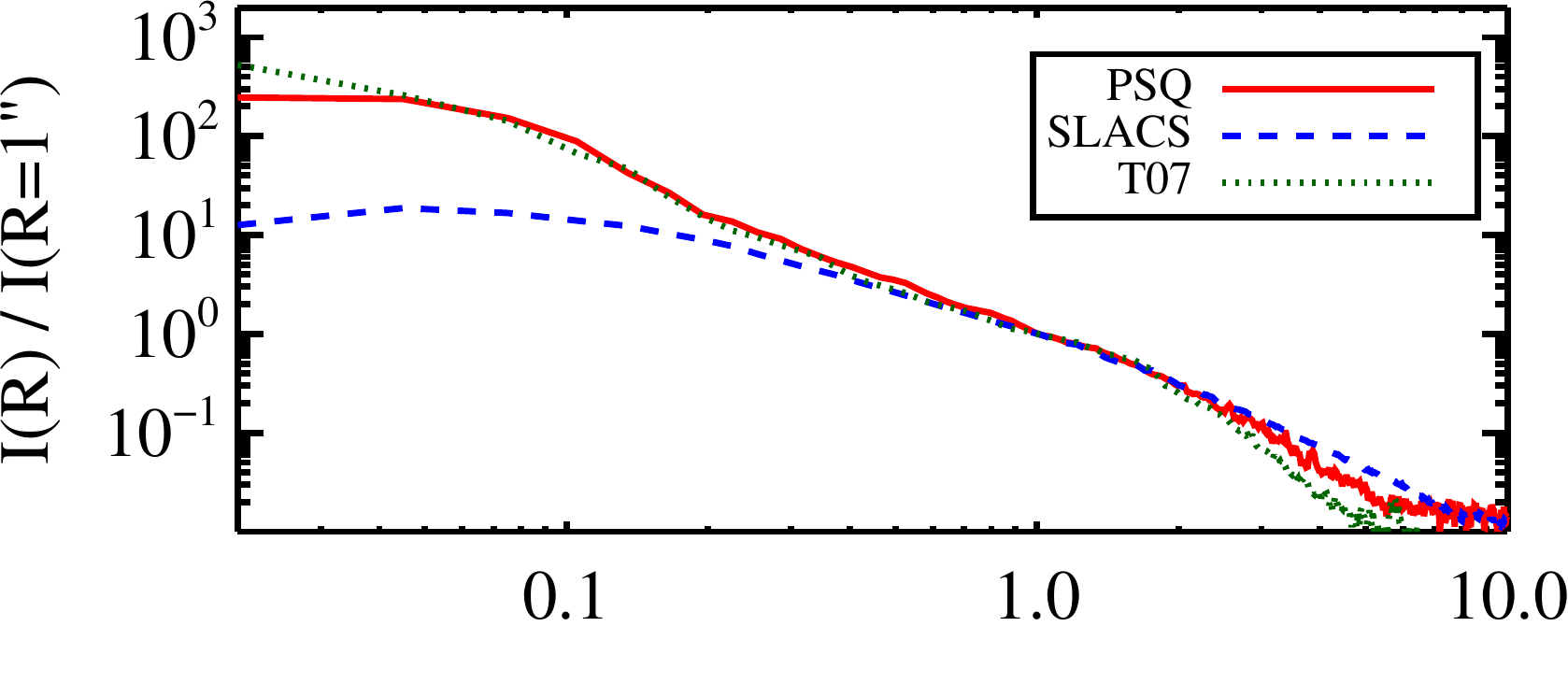}\vspace{-0.5cm}
\includegraphics[width=0.47\textwidth]{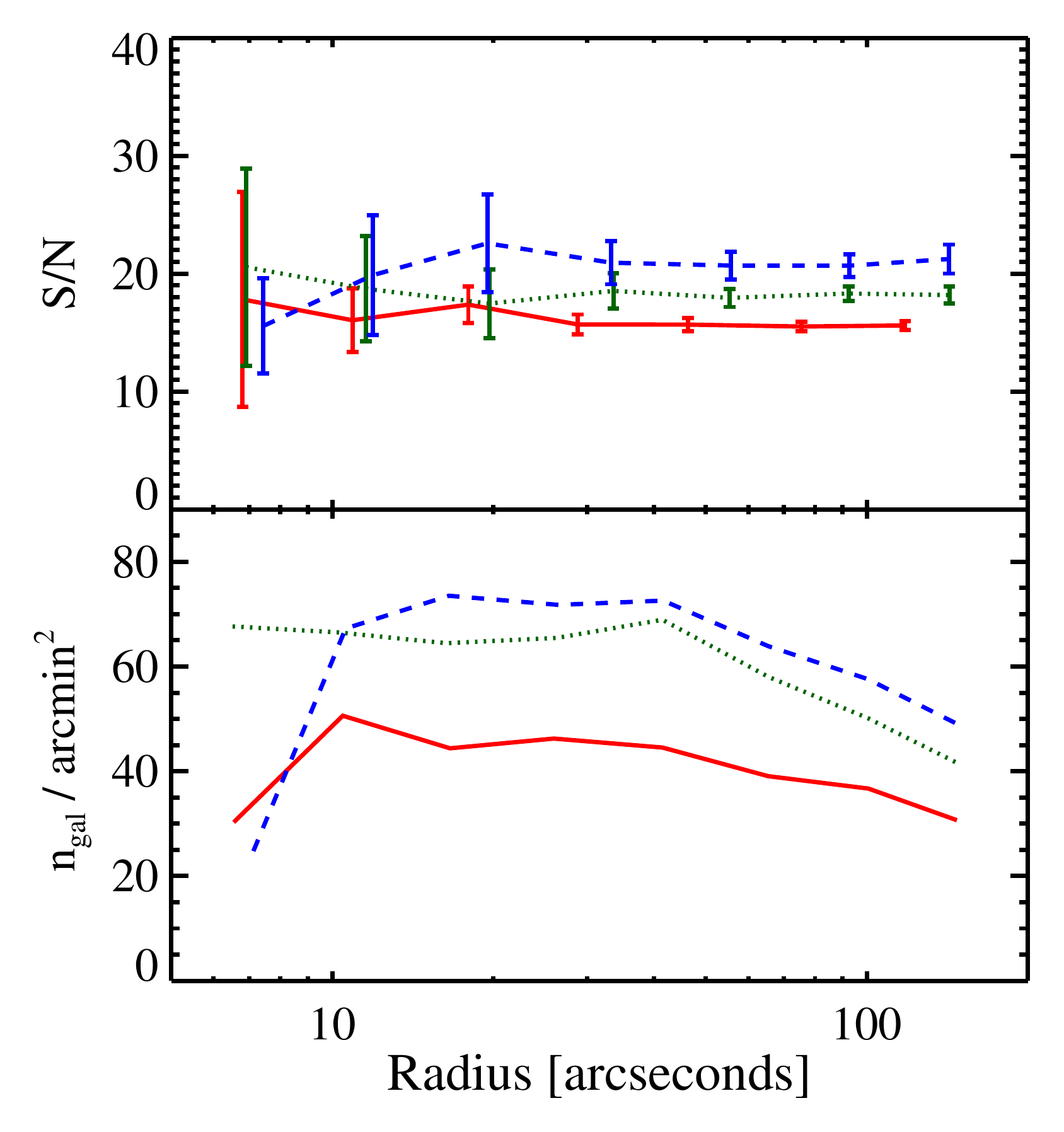}
\caption{ Systematic tests on the different samples of galaxies. We show in the top panel, ranging from 0.01 to 10 arc-seconds, the light profiles of the three different samples. The two quasar samples are comparable, and the contribution from the quasars can be clearly seen. The middle panel shows the median signal-to-noise for each source galaxy used in the analysis, and the bottom panel shows the source galaxy density as a function of distance from the centre of the galaxy. The middle and bottom panels span the entire field of view from 7 - 100 arc-seconds.}
\label{fig:all_tests}
\end{center}
\end{figure}

In addition to the cross-shear we also carried out checks to test for bias due to lens light, source galaxy redshift assumptions, noise bias and source galaxy blending.
We first checked that the presence of lens light was not a significant issue with our lensing estimation. To do this we measured the median lens light profile for each sample and normalized with respect to the intensity at a radius of 1 arc-second (where $I(R$=1$\arcsec)_{\rm PSQ} = 0.046$ e$^{-}$/second, $I(R$=1$\arcsec)_{\rm T07} = 0.028$e$^{-}$/second,  $I(R$=1$\arcsec)_{\rm SLACS} = 0.17$e$^{-}$/second). The top panel of Figure \ref{fig:all_tests} shows that the two quasar samples, PSQ and T07 both dropped to the sky background by $\sim4\arcsec$ or $\sim15$kpc and hence did not impact our shape measurement. The SLACS lenses were slightly larger, extending out to $\sim8\arcsec$, which could have impacted our source selection and shape measurement. To ensure that this was not the case we carried out two further checks. We first measured the signal-to-noise ratio (S/N) over the field for each sample, checking to see if sources were detected within the lens (and hence have unreliable shapes). Such an effect would have manifested itself as a mean increase in source galaxy S/N. We also examined the number density of galaxies over the field of view, checking to see if the source selection had picked up substructure within the lens that could mimic source galaxies. The bottom two panels of Figure \ref{fig:all_tests} show the results. We found that the S/N was constant across the field for all samples suggesting a robust source selection, and we also reported a constant number density of galaxies for the two quasar samples, also suggesting we did not select substructure within the lens. As expected the number density for the SLACS lenses drop in the inner most bin, this is due to an increase in noise due to foreground larger SLACS lenses. In addition to this we found that the innermost bin of the cross shear for the SLACS data was consistent with zero, further supporting the robustness of the source selection and shape measurement. The S/N also shows that the galaxies selected had a mean S/N$>15$ for all samples, where noise bias is negligible \citep{shape_bias}. 

The bottom two panels of Figure \ref{fig:all_tests} show the inter-sample variation in the number density and S/N.  This suggested that the mean source redshift also varied between samples. Since we assumed that all sources were at a redshift $z=1.0$  we tested the impact of this assumption on our results. We re-ran the analysis with a variety of different assumed source redshifts and studied the effect on our results. We note here that changing the source redshift would not affect the profile shape of the measured galaxies but only the mass estimates as the source redshift only affects $\Sigma_{\rm crit}$ in equation \eqref{eqn:crit}. However, despite this, we found that if we varied the source redshift from $z=0.8:1.5$, the change in derived masses were negligible compared to the statistical error bars. We therefore ignored this source of error. We finally checked to see if results were robust to source galaxy blending. In the method we removed all pairs of galaxies that were separated by less than $0.5\arcsec$, to check that our results did not depend on this we removed all galaxies that were separated by less than $1\arcsec$ and we repeated the experiment with an angular separation of $1\arcsec$, with the same results. We found no significant change in the profile of the galaxies and the derived masses, and were therefore confident in the source selection.

\subsection{Future surveys}
\begin{figure}
\begin{center}
\includegraphics[width=0.47\textwidth]{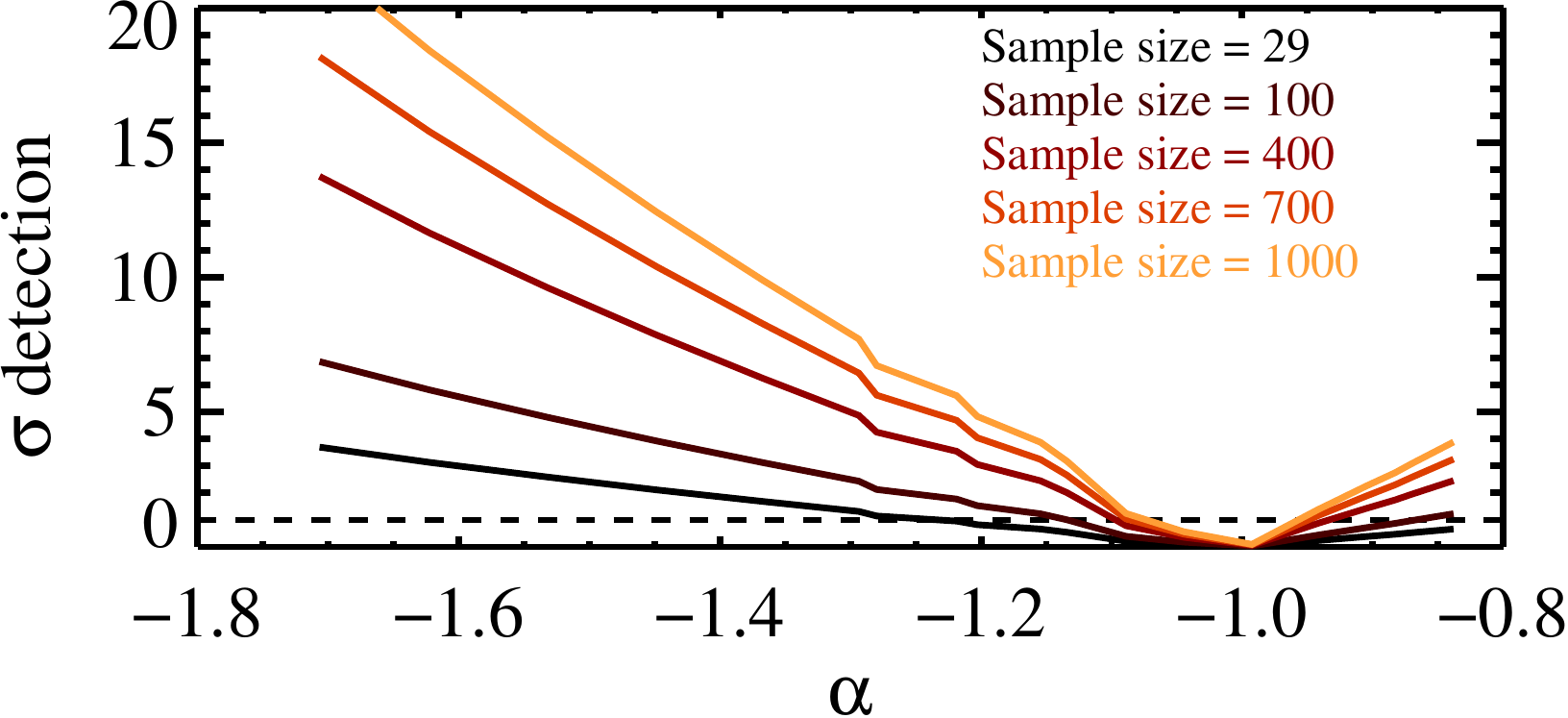}
\caption{ The sensitivity of our data to a change in profile. We artificially steepened the data to create a fake `steeper' profile, and fit a power law profile, measuring the index $\alpha$, and the associated error. We found that for our sample size of 29 quasars, we would measure a $1\sigma$ deviation from isothermality if the profile had exhibited an index of $\alpha\sim-1.2$ (where $\alpha_{\rm iso} = -1$). Future surveys will observe orders of magnitude more quasars resulting in more sensitive data. The dotted black line shows a detected $1\sigma$ departure from isothermality.}
\label{fig:steep}
\end{center}
\end{figure}

Our study of 29 PSQs showed that the dark matter halo in these galaxies did not depart from isothermality. Given the accuracy of our measurements we tested how sensitive our measurements were to a change in profile, and how much it would need to change in order for us to detect a discrepancy. Additionally, with the advent of all sky space surveys we investigated how sensitive future measurements are expected to be. To do this we fitted a very simple power law profile to the data, $\Delta\Sigma \propto r^\alpha$ and measured the index, $\alpha$ and its associated error. We then artificially steepened the data, raising the index and refitted the profile, measuring the index with an associated error. We then compared this to an isothermal profile ( $\alpha = -1$) and determined how significant the departure from isothermality we had measured. As the data became steeper we were able to better detect a departure from isothermality. Figure \ref{fig:steep} shows the relation between the measured gradient of the profile and the measured significance of a departure from isothermality. We repeated this exercise using different fake sample sizes, decreasing the error bars by a factor $\sqrt{N_{\rm quasars}}$. We found that with future surveys we will be able to detect a 10\% departure from isothermality at the $1\sigma$ confidence level using $\sim1000$ quasars. Such a sample size will be possible with surveys such as Euclid \citep{EUCLID} in the near future.

\section{Conclusions}
We measured the total mass profile based solely on post starburst quasars (PSQ) host galaxies that showed signs of recent merger activity. Our measurement, based on weak gravitational lensing of background galaxies, is compared with that of the  SLACS sample of massive elliptical galaxies and with a sample of luminous Seyfert galaxies. We found that in all three cases the profiles did not depart significantly from isothermallity and concluded that within the error bars shown, mergers do not affect the shape of the dark matter halo profile. 
Given the commonality of the two quasar sample in NFW mass, de Vaucouleur mass, and central black hole mass, we found that AGN activity did not affect the shape of the dark matter profile, and therefore in these two samples of galaxies baryonic feedback was not strong enough to affect the potential via gravitational or non-gravitational forces.

This study has presented the first total mass and mass profile measurement of non-obscured, low redshift quasars that show recent signs of merging. Given the consistency of the results and the small sample size we tested to see how sensitive the data actually was to a change in profile shape. We found with the current data we would be able to detect a $\sim20\%$ departure from isothermality at the $1\sigma$ level, however with future surveys we will have access to many more quasars, reducing this sensitivity to a $>10$\% change in profile shape. However, these estimates are purely statistical and the robustness of these results would need to be tested in the event of more accurate data.

\section*{Acknowledgements}
We would like to thank F. Combes and Y. Revaz for helpful discussions. We would also like to thank R. Massey and J. Rhodes for help on HST data reduction and shape measurement. The authors are supported by the Swiss National Science Foundation (SNSF). This research is fully based on data taken from the HST archive, which we warmly acknowledge. The results here are based on observations made with the NASA/ESA Hubble Space Telescope, obtained from the Data Archive at the Space Telescope Science Institute, which is operated by the Association of Universities for Research in Astronomy, Inc., under NASA contract NAS 5-26555. These observations are associated with program \#10216,10588,10494 and 10886.

\bibliographystyle{mn2e}
\bibliography{bibliography}

\begin{thebibliography}{52}
\expandafter\ifx\csname natexlab\endcsname\relax\def\natexlab#1{#1}\fi

\bibitem[{{Amara} \& {R{\'e}fr{\'e}gier}(2008)}]{shape_bias}
{Amara} A., {R{\'e}fr{\'e}gier} A., 2008, \mnras, 391, 228

\bibitem[{{Auger}(2008)}]{Auger2008}
{Auger} M.~W., 2008, \mnras, 383, L40

\bibitem[{{Bahcall} {et~al}\mbox{.}(1997){Bahcall}, {Kirhakos}, {Saxe}, \&
  {Schneider}}]{Bahcall1997}
{Bahcall} J.~N., {Kirhakos} S., {Saxe} D.~H., {Schneider} D.~P., 1997, \apj,
  479, 642

\bibitem[{{Barger} {et~al}\mbox{.}(2005){Barger}, {Cowie}, {Mushotzky}, {Yang},
  {Wang}, {Steffen}, \& {Capak}}]{downsizing}
{Barger} A.~J., {Cowie} L.~L., {Mushotzky} R.~F., {Yang} Y., {Wang} W.-H.,
  {Steffen} A.~T., {Capak} P., 2005, \aj, 129, 578

\bibitem[{{Barger} {et~al}\mbox{.}(2014){Barger}, {Cowie}, {Owen}, {Chen},
  {Hasinger}, {Hsu}, \& Y.}]{BHstarform}
{Barger} A.~J., {Cowie} L.~L., {Owen} F.~N., {Chen} C.-C., {Hasinger} G., {Hsu}
  L.-Y., Y. L., 2014, "arXiv"

\bibitem[{{Barnab{\`e}} {et~al}\mbox{.}(2011){Barnab{\`e}}, {Czoske},
  {Koopmans}, {Treu}, \& {Bolton}}]{Barnabe2011}
{Barnab{\`e}} M., {Czoske} O., {Koopmans} L.~V.~E., {Treu} T., {Bolton} A.~S.,
  2011, \mnras, 415, 2215

\bibitem[{{Bartelmann}(1996)}]{proj_NFW}
{Bartelmann} M., 1996, \aap, 313, 697

\bibitem[{{Bolton} {et~al}\mbox{.}(2006){Bolton}, {Burles}, {Koopmans}, {Treu},
  \& {Moustakas}}]{SLACS}
{Bolton} A.~S., {Burles} S., {Koopmans} L.~V.~E., {Treu} T., {Moustakas} L.~A.,
  2006, \apj, 638, 703

\bibitem[{{Brainerd} {et~al}\mbox{.}(1996){Brainerd}, {Blandford}, \&
  {Smail}}]{gal_gal}
{Brainerd} T.~G., {Blandford} R.~D., {Smail} I., 1996, \apj, 466, 623

\bibitem[{{Cales} {et~al}\mbox{.}(2011){Cales}, {Brotherton}, {Shang},
  {Bennert}, {Canalizo}, {Stoll}, {Ganguly}, {Vanden Berk}, {Paul}, \&
  {Diamond-Stanic}}]{PSQ1}
{Cales} S.~L. {et~al.}, 2011, \apj, 741, 106

\bibitem[{{Cales} {et~al}\mbox{.}(2013){Cales}, {Brotherton}, {Shang},
  {Runnoe}, {DiPompeo}, {Bennert}, {Canalizo}, {Hiner}, {Stoll}, {Ganguly}, \&
  {Diamond-Stanic}}]{PSQ2}
{Cales} S.~L. {et~al.}, 2013, \apj, 762, 90

\bibitem[{{de Vaucouleurs}(1948)}]{vauc}
{de Vaucouleurs} G., 1948, Annales d'Astrophysique, 11, 247

\bibitem[{{Di Matteo} {et~al}\mbox{.}(2005){Di Matteo}, {Springel}, \&
  {Hernquist}}]{merger_qso}
{Di Matteo} T., {Springel} V., {Hernquist} L., 2005, \nat, 433, 604

\bibitem[{{Disney} {et~al}\mbox{.}(1995){Disney}, {Boyce}, {Blades},
  {Boksenberg}, {Crane}, {Deharveng}, {Macchetto}, {Mackay}, {Sparks}, \&
  {Phillipps}}]{disney95}
{Disney} M.~J. {et~al.}, 1995, \nat, 376, 150

\bibitem[{{Dobke} {et~al}\mbox{.}(2007){Dobke}, {King}, \&
  {Fellhauer}}]{Dobke2007}
{Dobke} B.~M., {King} L.~J., {Fellhauer} M., 2007, \mnras, 377, 1503

\bibitem[{{Dunlop} {et~al}\mbox{.}(2003){Dunlop}, {McLure}, {Kukula}, {Baum},
  {O'Dea}, \& {Hughes}}]{dunlop03}
{Dunlop} J.~S., {McLure} R.~J., {Kukula} M.~J., {Baum} S.~A., {O'Dea} C.~P.,
  {Hughes} D.~H., 2003, \mnras, 340, 1095

\bibitem[{{Emsellem} {et~al}\mbox{.}(2014){Emsellem}, {Renaud}, {Bournaud},
  {Elmegreen}, {Combes}, \& {Gabor}}]{BH_gal_form2}
{Emsellem} E., {Renaud} F., {Bournaud} F., {Elmegreen} B., {Combes} F., {Gabor}
  J., 2014, ArXiv e-prints

\bibitem[{{Englmaier} \& {Shlosman}(2000)}]{BH_gal_form}
{Englmaier} P., {Shlosman} I., 2000, \apj, 528, 677

\bibitem[{{Ferrarese} \& {Merritt}(2000)}]{BH_bulge}
{Ferrarese} L., {Merritt} D., 2000, \apjl, 539, L9

\bibitem[{{Gavazzi} {et~al}\mbox{.}(2007){Gavazzi}, {Treu}, {Rhodes},
  {Koopmans}, {Bolton}, {Burles}, {Massey}, \& {Moustakas}}]{SLACS_SL}
{Gavazzi} R., {Treu} T., {Rhodes} J.~D., {Koopmans} L.~V.~E., {Bolton} A.~S.,
  {Burles} S., {Massey} R.~J., {Moustakas} L.~A., 2007, \apj, 667, 176

\bibitem[{{Gebhardt} {et~al}\mbox{.}(2000){Gebhardt}, {Bender}, {Bower},
  {Dressler}, {Faber}, {Filippenko}, {Green}, {Grillmair}, {Ho}, {Kormendy},
  {Lauer}, {Magorrian}, {Pinkney}, {Richstone}, \& {Tremaine}}]{BH_bulge2}
{Gebhardt} K. {et~al.}, 2000, \apjl, 539, L13

\bibitem[{{Harvey} {et~al}\mbox{.}(2015){Harvey}, {Massey}, {Kitching},
  {Taylor}, \& {Tittley}}]{Harvey15}
{Harvey} D., {Massey} R., {Kitching} T., {Taylor} A., {Tittley} E., 2015,
  Science, 347, 1462

\bibitem[{{Hasinger} {et~al}\mbox{.}(2005){Hasinger}, {Miyaji}, \&
  {Schmidt}}]{downsizing1}
{Hasinger} G., {Miyaji} T., {Schmidt} M., 2005, \aap, 441, 417

\bibitem[{{Heckman} {et~al}\mbox{.}(2004){Heckman}, {Kauffmann}, {Brinchmann},
  {Charlot}, {Tremonti}, \& {White}}]{downsize}
{Heckman} T.~M., {Kauffmann} G., {Brinchmann} J., {Charlot} S., {Tremonti} C.,
  {White} S.~D.~M., 2004, \apj, 613, 109

\bibitem[{{Heymans} {et~al}\mbox{.}(2012){Heymans}, {Van Waerbeke}, {Miller},
  {Erben}, {Hildebrandt}, {Hoekstra}, {Kitching}, {Mellier}, {Simon},
  {Bonnett}, {Coupon}, {Fu}, {Harnois D{\'e}raps}, {Hudson}, {Kilbinger},
  {Kuijken}, {Rowe}, {Schrabback}, {Semboloni}, {van Uitert}, {Vafaei}, \&
  {Velander}}]{CHFTLenS}
{Heymans} C. {et~al.}, 2012, \mnras, 427, 146

\bibitem[{{Hoekstra} {et~al}\mbox{.}(2005){Hoekstra}, {Hsieh}, {Yee}, {Lin}, \&
  {Gladders}}]{gal_gal_isolated}
{Hoekstra} H., {Hsieh} B.~C., {Yee} H.~K.~C., {Lin} H., {Gladders} M.~D., 2005,
  \apj, 635, 73

\bibitem[{{Hoekstra} {et~al}\mbox{.}(2004){Hoekstra}, {Yee}, \&
  {Gladders}}]{gal_gal_1halo2}
{Hoekstra} H., {Yee} H.~K.~C., {Gladders} M.~D., 2004, \apj, 606, 67

\bibitem[{{Hopkins} \& {Hernquist}(2006)}]{accretion_qso}
{Hopkins} P.~F., {Hernquist} L., 2006, \apjs, 166, 1

\bibitem[{{Hopkins} \& {Hernquist}(2009)}]{accretion_qso1}
{Hopkins} P.~F., {Hernquist} L., 2009, \apj, 694, 599

\bibitem[{{Hopkins} {et~al}\mbox{.}(2006){Hopkins}, {Hernquist}, {Cox}, {Di
  Matteo}, {Robertson}, \& {Springel}}]{merger_qso2}
{Hopkins} P.~F., {Hernquist} L., {Cox} T.~J., {Di Matteo} T., {Robertson} B.,
  {Springel} V., 2006, \apjs, 163, 1

\bibitem[{{Hudson} {et~al}\mbox{.}(1998){Hudson}, {Gwyn}, {Dahle}, \&
  {Kaiser}}]{gal_gal_1halo1}
{Hudson} M.~J., {Gwyn} S.~D.~J., {Dahle} H., {Kaiser} N., 1998, \apj, 503, 531

\bibitem[{{Jahnke} {et~al}\mbox{.}(2007){Jahnke}, {Wisotzki}, {Courbin}, \&
  {Letawe}}]{Jahnke2007}
{Jahnke} K., {Wisotzki} L., {Courbin} F., {Letawe} G., 2007, \mnras, 378, 23

\bibitem[{{Laureijs} {et~al}\mbox{.}(2011){Laureijs}, {Amiaux}, {Arduini},
  {Augu{\`e}res}, {Brinchmann}, {Cole}, {Cropper}, {Dabin}, {Duvet}, {Ealet},
  \& et~al.}]{EUCLID}
{Laureijs} R. {et~al.}, 2011, arXiv:1110.3193

\bibitem[{{Leauthaud} {et~al}\mbox{.}(2014){Leauthaud}, {Benson}, {Civano},
  {Coil}, {Bundy}, {Massey}, {Schramm}, {Schulze}, {Capak}, {Elvis}, {Kulier},
  \& {Rhodes}}]{cosmos_quasar}
{Leauthaud} A. {et~al.}, 2014, ArXiv e-prints

\bibitem[{{Leauthaud} {et~al}\mbox{.}(2007){Leauthaud}, {Massey}, {Kneib},
  {Rhodes}, {Johnston}, {Capak}, {Heymans}, {Ellis}, {Koekemoer}, {Le
  F{\`e}vre}, {Mellier}, {R{\'e}fr{\'e}gier}, {Robin}, {Scoville}, {Tasca},
  {Taylor}, \& {Van Waerbeke}}]{COSMOSintdisp}
{Leauthaud} A. {et~al.}, 2007, \apjs, 172, 219

\bibitem[{{Leauthaud} {et~al}\mbox{.}(2012){Leauthaud}, {Tinker}, {Bundy},
  {Behroozi}, {Massey}, {Rhodes}, {George}, {Kneib}, {Benson}, {Wechsler},
  {Busha}, {Capak}, {Cort{\^e}s}, {Ilbert}, {Koekemoer}, {Le F{\`e}vre},
  {Lilly}, {McCracken}, {Salvato}, {Schrabback}, {Scoville}, {Smith}, \&
  {Taylor}}]{gal_gal_2halo2}
{Leauthaud} A. {et~al.}, 2012, \apj, 744, 159

\bibitem[{{Letawe} {et~al}\mbox{.}(2007){Letawe}, {Magain}, {Courbin},
  {Jablonka}, {Jahnke}, {Meylan}, \& {Wisotzki}}]{Letawe2007}
{Letawe} G., {Magain} P., {Courbin} F., {Jablonka} P., {Jahnke} K., {Meylan}
  G., {Wisotzki} L., 2007, \mnras, 378, 83

\bibitem[{{Letawe} {et~al}\mbox{.}(2008){Letawe}, {Magain}, {Letawe},
  {Courbin}, \& {Hutsem{\'e}kers}}]{Letawe2008}
{Letawe} Y., {Magain} P., {Letawe} G., {Courbin} F., {Hutsem{\'e}kers} D.,
  2008, \apj, 679, 967

\bibitem[{{Mandelbaum} {et~al}\mbox{.}(2009){Mandelbaum}, {Li}, {Kauffmann}, \&
  {White}}]{agn_gal_gal}
{Mandelbaum} R., {Li} C., {Kauffmann} G., {White} S.~D.~M., 2009, \mnras, 393,
  377

\bibitem[{{Mandelbaum} {et~al}\mbox{.}(2006){Mandelbaum}, {Seljak},
  {Kauffmann}, {Hirata}, \& {Brinkmann}}]{gal_gal_2halo}
{Mandelbaum} R., {Seljak} U., {Kauffmann} G., {Hirata} C.~M., {Brinkmann} J.,
  2006, \mnras, 368, 715

\bibitem[{{Maoz} \& {Rix}(1993)}]{vauc1}
{Maoz} D., {Rix} H.-W., 1993, \apj, 416, 425

\bibitem[{{Massey}(2010)}]{CTI}
{Massey} R., 2010, \mnras, 409, L109

\bibitem[{{Massey} {et~al}\mbox{.}(2014){Massey}, {Schrabback}, {Cordes},
  {Marggraf}, {Israel}, {Miller}, {Hall}, {Cropper}, {Prod'homme}, \& {Matias
  Niemi}}]{CTI2}
{Massey} R. {et~al.}, 2014, \mnras, 439, 887

\bibitem[{{McLure} {et~al}\mbox{.}(1999){McLure}, {Kukula}, {Dunlop}, {Baum},
  {O'Dea}, \& {Hughes}}]{mclure99}
{McLure} R.~J., {Kukula} M.~J., {Dunlop} J.~S., {Baum} S.~A., {O'Dea} C.~P.,
  {Hughes} D.~H., 1999, \mnras, 308, 377

\bibitem[{{Navarro} {et~al}\mbox{.}(1997){Navarro}, {Frenk}, \& {White}}]{NFW}
{Navarro} J.~F., {Frenk} C.~S., {White} S.~D.~M., 1997, \apj, 490, 493

\bibitem[{{Rhodes} {et~al}\mbox{.}(2000){Rhodes}, {Refregier}, \&
  {Groth}}]{RRG}
{Rhodes} J., {Refregier} A., {Groth} E.~J., 2000, \apj, 536, 79

\bibitem[{{Schneider} \& {Rix}(1997)}]{gal_gal_1halo}
{Schneider} P., {Rix} H.-W., 1997, \apj, 474, 25

\bibitem[{{Springel} {et~al}\mbox{.}(2005){Springel}, {Di Matteo}, \&
  {Hernquist}}]{merger_qso1}
{Springel} V., {Di Matteo} T., {Hernquist} L., 2005, \apjl, 620, L79

\bibitem[{{Tremaine} {et~al}\mbox{.}(2002){Tremaine}, {Gebhardt}, {Bender},
  {Bower}, {Dressler}, {Faber}, {Filippenko}, {Green}, {Grillmair}, {Ho},
  {Kormendy}, {Lauer}, {Magorrian}, {Pinkney}, \& {Richstone}}]{bhvelrel}
{Tremaine} S. {et~al.}, 2002, \apj, 574, 740

\bibitem[{{Treu} {et~al}\mbox{.}(2007){Treu}, {Woo}, {Malkan}, \&
  {Blandford}}]{coevolutuionBH}
{Treu} T., {Woo} J.-H., {Malkan} M.~A., {Blandford} R.~D., 2007, \apj, 667, 117

\bibitem[{{van Uitert} {et~al}\mbox{.}(2011){van Uitert}, {Hoekstra},
  {Velander}, {Gilbank}, {Gladders}, \& {Yee}}]{gal_gal_2halo1}
{van Uitert} E., {Hoekstra} H., {Velander} M., {Gilbank} D.~G., {Gladders}
  M.~D., {Yee} H.~K.~C., 2011, \aap, 534, A14

\bibitem[{{Velander} {et~al}\mbox{.}(2014){Velander}, {van Uitert}, {Hoekstra},
  {Coupon}, {Erben}, {Heymans}, {Hildebrandt}, {Kitching}, {Mellier}, {Miller},
  {Van Waerbeke}, {Bonnett}, {Fu}, {Giodini}, {Hudson}, {Kuijken}, {Rowe},
  {Schrabback}, \& {Semboloni}}]{chft_gal_gal}
{Velander} M. {et~al.}, 2014, \mnras, 437, 2111

\end{thebibliography}

\bsp
\label{lastpage}

\end{document}